\documentclass[12pt]{article}
\newcommand{\ald}{\dot \alpha}

\newcommand{\beq}{\begin{equation}}
\newcommand{\eeq}{\end{equation}}
\begin{document}
\title{Gravitationally dressed Parke-Taylor amplitudes.}
\author{K.G.Selivanov}
\date{ITEP-TH-64/97}
\maketitle

\begin{abstract}
A generating function for the Parke-Taylor amplitudes with any number
of positive helicity gravitons in addition to the positive helicity gluons
is obtained using the recently constructed self-dual classical solution
of the type of perturbiner in Yang-Mills theory interacting with gravity.
\end{abstract}

Computation of multi-particle amplitudes in gauge theories in the leading
and next-to-leading orders has turned into the whole industry. There is
permanent technological progress in the canonical Feynman perturbation theory
direction (see, e.g., the reviews \cite{mapa}, \cite{dixon}) and there are also 
alternative directions like the string inspired technique (see, e.g. the review
\cite{BK}) and the world-line technique (\cite{St}). Our favorite approach
is based on the so-called perturbiners (ptb). 

Ptbs are solutions of field 
equations which are generating functions for tree form-factors in the theory.
The ptb solution can be given an intrinsic definition \cite{RS1} which is formally
independent of the Feynman diagrams. In words it sounds as follows.
Take a solution of the free (linear) field equation in the theory under consideration 
in the form of superposition of a set of plane waves. Regard every plane
wave as nilpotent. The corresponding ptb is a solution of the full (nonlinear) field 
equation which is 
polynomial in the plane waves, first order term of the polynomial being just
the solution of the free field equation. The plane waves are nothing but the asymptotic 
wave functions of the one-particle states included in the form-factors which the
ptb is the generating function for. The nilpotency assumes that the ptb is the generating 
function for form-factors without identical one-particle states. Obviously, there is no 
a loss of generality in the assumption of nilpotency.

In gauge theories, when one reduces the set of asymptotic states to the positive
helicity sector, one can consider only self-dual (SD)  ptb solution (this fact was pointed
out in \cite{Ba} and, independently, in \cite{Se} for the pure Yang-Mills (YM) case).
Thus solving the SD equation one obtains the generating function for the tree form-factor
in the same helicity sector of the theory. In the pure YM case these form-factors, 
called also off-shell currents, had already existed in the literature before obtaining
them via solution of the SD equation. They were obtained in \cite{BG} as solutions
for recursion relations deduced from the Feynman diagrams. Actually, in \cite{Ba}
it was first shown that the recursion relations \cite{BG} follow from the SD equation
and the corresponding solution of the SD equation was obtained in terms of solutions of 
the recursion relations obtained previously in \cite{BG}. A similar SD solution was 
discussed in \cite{KO} and their consideration was also based on recursion relations
analogous to \cite{BG}. In \cite{RS2} we obtained the SD ptb in YM theory by means
of the twistor construction \cite{Ward}. Miraculously, the tree form-factors in the 
positive helicity sector of YM theory appeared to be expressed in terms of a meromorphic
function on an auxiliary 2d sphere with values in (the complexification of) the gauge 
group. Locations of poles of the meromorphic function were defined by momenta
of the gluons and the residues were, roughly speaking, equal to their asymptotic 
wave functions.

In \cite{RS3} we obtained the SD ptb solution in gravity and in \cite{Se1} the SD ptb 
was obtained in YM+gravity case. Again, the twistor methods \cite{Penrose}, \cite{AHS}
happened to be extremely efficient. Up to our best knowledge, the tree positive
helicity form-factors in gravity and in YM+gravity cases had not existed in the literature
before \cite{RS3} and \cite{Se1}.

Once the SD ptb is known one can develop perturbation theory in two directions:
to include loops and to include the opposite helicity particles. We did not touch 
the first  direction yet, as far as concerned to the second one, in \cite{RS2} we
solved the linearization of YM equations in the SD ptb background and thus obtained
a generating function for the amplitudes with two gluons in the opposite helicity
sector, so-called the maximally helicity violating amplitudes or the Parke-Taylor 
amplitudes \cite{PT}.

In this letter I report on an analogous work in the YM+gravity  case. I solve the 
linearization of YM equations in the background of the SD ptb solution in YM+gravity
and thus I obtain generalization of the Parke-Taylor amplitudes to include any number
of the positive helicity gravitons. It seems that the traditional Feynman diagrams
technique would be of no help in computation of these amplitudes. The string inspired
consideration in a bit different from ours context of pure gravity led the authors of 
\cite{gravPT} to conjectured expressions for maximally helicity violating amplitudes
which are yet to be proved or disproved.

Before presenting the results let us introduce some notation. 
$k_{l}^{\alpha \ald}, \alpha=1,2, \ald=\dot 1, \dot 2, l=1, \ldots , j$ will stand for 
four-momenta of the positive helicity gluons (we use the spinor notations).
$k'^{\alpha \ald}$ and $k''^{\alpha \ald}$ will stand for four-momenta of the two
negative helicity gluons. Since  $k_{l}^{\alpha \ald}$ is a light-like four-vector
it decomposes into a product of two spinors,  
$k_{l}^{\alpha \ald}={\ae}_{l}^{\alpha}{\lambda}_{l}^{\ald}$.{\footnote
{the reality of the four momentum in Minkowski space assumes that $\lambda^{\ald}=
{\bar {\ae}}^{\alpha}$}} Analogously, $s_{n}^{\alpha \ald},  
\alpha=1,2, \ald=\dot 1, \dot 2, n{\in}N$ will stand for gravitons four-momenta.
Again, 
$s_{n}^{\alpha \ald}={\sigma}_{n}^{\alpha}{\rho}_{n}^{\ald}$. The notation
$({\ae}_{i}{\ae}_{j})$ will stand for the contraction of the spinors with the 
standard antisymmetric $\epsilon$-symbol, 
$({\ae}_{i}{\ae}_{j})={\ae}_{i \alpha}{\ae}_{j}^{\alpha}=
{\ae}_{i \alpha}{\epsilon}^{\alpha \beta}{\ae}_{j \beta}$ (and analogously for the 
dotted indices).

Our results for the gravitationally dressed Parke-Taylor amplitudes can be expressed
by the following generating function
\begin{eqnarray}
\label{one}
M('',1,2, \ldots , l,',l+1, \ldots , j, \{s_{n}\}, \{a_{n}\}, n{\in}N)=\nonumber\\
i\frac{({\ae}''{\ae}')^{4}\int d^{4}x e^{(k''+k')x} 
\prod_{l=1}^{j}E_{l}(x, \{s_{n}\}, \{a_{n}\})}
{({\ae}''{\ae}_{1})({\ae}_{1}{\ae}_{2}) \ldots ({\ae}_{l-1}{\ae}_{l})
({\ae}_{l}{\ae}')({\ae}'{\ae}_{l+1}) \ldots ({\ae}_{j-1}{\ae}_{j})
({\ae}_{j}{\ae}'')}
\end{eqnarray}
where\\
$E_{l}(x, \{s_{n}\}, \{a_{n}\})=e^{k_{l}x+\sum_{n_{1}}a_{n_{1}}
\frac{({\ae}_{l}{\ae}')({\ae}_{l}{\ae}'')(\lambda_{l}\rho_{n_{1}})}
{({\sigma}_{n_{1}}{\ae}')({\sigma}_{n_{1}}{\ae}'')({\ae}_{l}{\sigma}_{n_{1}})}
e^{s_{n_{1}}x+\sum_{n_{2}}a_{n_{2}}
\frac{({\sigma}_{n_{1}}{\ae}')({\sigma}_{n_{1}}{\ae}'')(\rho_{n_{1}}\rho_{n_{2}})}
{({\sigma}_{n_{2}}{\ae}')({\sigma}_{n_{2}}{\ae}'')
({\sigma}_{n_{1}}{\sigma}_{n_{2}})}e^{\ldots}}}$\\
$a_{n}$ are symbols of gravitonic annihilation/creation operators, they are assumed 
to be nilpotent, $a_{n}^{2}=0$ (as explained above, the nilpotency assumes that 
$M$ is the generating function for the amplitudes without identical gravitonic states). 
$M$ from Eq.(\ref{one}) is the generating function for the cyclically color ordered
amplitudes, the cyclic color order is assumed to be
$('',1,2, \ldots , l,',l+1, \ldots , j)$. The amplitudes arise as coefficients in the Taylor
expansion of $M$  in powers of $a$'s. $\int d^{4}x$-integration just produces
the concervation law $\delta$-function in every amplitudes.
As a particular case of Eq.(\ref{one}), one can see that the amplitudes of scattering
of two gluons (photons) of any helicities into any number of the positive helicity 
gravitons are zero. 

Derivation of the results presented will be described elsewhere. 

It's my pleasure to thank A.Rosly for discussions. The work was partially
supported by INTAS-96-482.

\end{document}